\documentclass[paper=A4,twoside,11pt,DIV=12,headings=big]{scrartcl}
\pdfoutput=1

\usepackage[a4paper, top=1.2in, bottom=1.2in, left=1.1in, right=1.1in]{geometry}
\usepackage{amsfonts}
\usepackage{amsmath,amssymb}
\usepackage{graphicx,hyperref}

\newcommand{\mc}{\mathcal}

\newcommand{\mumu}{\mu^+ \mu^-}

\def\sla#1{\setbox0=\hbox{$#1$}\dimen0=\wd0
      \setbox1=\hbox{/} \dimen1=\wd1 \ifdim\dimen0>\dimen1
      \rlap{\hbox to \dimen0{\hfil/\hfil}} #1                        \else
      \rlap{\hbox to \dimen1{\hfil$#1$\hfil}}
      /   \fi}

\newcommand{\be}{\begin{equation}}
\newcommand{\ee}{\end{equation}}
\newcommand{\bea}{\begin{eqnarray}}
\newcommand{\eea}{\end{eqnarray}}

\newcommand{\nn}{\nonumber}

\DeclareOldFontCommand{\rm}{\normalfont\rmfamily}{\mathrm}
\DeclareOldFontCommand{\sf}{\normalfont\sffamily}{\mathsf}
\DeclareOldFontCommand{\tt}{\normalfont\ttfamily}{\mathtt}
\DeclareOldFontCommand{\bf}{\normalfont\bfseries}{\mathbf}
\DeclareOldFontCommand{\it}{\normalfont\itshape}{\mathit}
\DeclareOldFontCommand{\sl}{\normalfont\slshape}{\@nomath\sl}
\DeclareOldFontCommand{\sc}{\normalfont\scshape}{\@nomath\sc}

\begin{document}

\markboth{Diego Guadagnoli}{Flavour anomalies on the eve of the Run-2 verdict}

\thispagestyle{plain}

\noindent
{\small Preprint of invited short review published in\\MPLA, Volume 32, Issue 07, DOI: 10.1142/S0217732317300063 \copyright~[copyright World Scientific Publishing Company], {\tt www.worldscinet.com/mpla}}

\begin{flushright}
\small
LAPTH-004/17
\end{flushright}

\begin{center}
{\sffamily \bfseries \LARGE \boldmath
Flavour anomalies\\[0.2cm]on the eve of the Run-2 verdict}\\[0.8 cm]
{\normalsize \sffamily \bfseries Diego Guadagnoli} \\[0.5 cm]
\small
{\em Laboratoire d'Annecy-le-Vieux de Physique Th\'eorique UMR5108\,, \\CNRS et Universit\'e de Savoie Mont-Blanc,\\B.P.~110, Annecy-le-Vieux, F-74941 Annecy Cedex, France\\
[0.2cm]
diego.guadagnoli@lapth.cnrs.fr}
\end{center}

\begin{abstract}

\noindent The $R_K$ measurement by LHCb suggests non-standard lepton-universality violation (LUV) to occur in $b \to s \ell^+ \ell^-$ decays, with effects in muons rather than electrons. It is intriguing that a number of other measurements of $b \to s \ell^+ \ell^-$ transitions by LHCb and $B$-factories are consistent in magnitude and sign with the $R_K$ effect, and fit a coherent effective-theory picture. Further indications of non-standard LUV are provided by the long-standing discrepancies in $b \to c \tau \nu$ transitions via the ratios $R(D)$ and $R(D^*)$. We review in detail the experimental situation and its rich outlook, the theoretical efforts -- and their challenges -- towards convincing dynamics beyond the effective-theory level, and discuss the many directions of further investigation that propagate from the current situation.

\end{abstract}


\medskip


\bigskip

\section{Introduction}
\label{sec:intro}

\noindent A rather striking qualitative feature of the data collected so far at the LHCb experiment is the fact that a whole range of $b \to s$ data involving a $\mumu$ pair display a consistent pattern, with experimental data being below the respective Standard-Model (SM) prediction, for di-lepton invariant masses below the charmonium threshold. This is true for the $B^{0} \to K^{0} \mumu$, the $B^{+} \to K^{+} \mumu$ and the $B^{+} \to K^{*+} \mumu$ decays \cite{Aaij:2014pli}, for the $B^0_s \to \phi \mumu$ decay \cite{Aaij:2015esa} and, very recently, even in hyperon channels for the $\Lambda_b \to \Lambda \mumu$ decay \cite{Aaij:2015xza,Detmold:2016pkz}. We know that branching-ratio measurements suffer from large theoretical uncertainties, especially because of the poor knowledge of hadronic form factors. However, here is a clean quantity: the ratio $R_K$ \cite{Aaij:2014ora}
\bea
\label{eq:RK}
R_K ~\equiv~ \frac{\mc B (B^+ \to K^+ \mu^+\mu^-)}{\mc B (B^+ \to K^+ e^+e^-)} ~=~
0.745^{+0.090}_{-0.074}\,{\rm (stat)}\pm 0.036\,{\rm (syst)}\,,
\eea
as measured in the di-lepton invariant-mass-squared range $[1, 6]$ GeV$^{2}$. The SM predicts unity with percent-level corrections \cite{Bordone:2016gaq,Bobeth:2007dw,Bouchard:2013mia,Hiller:2003js}, implying a $2.6\sigma$ discrepancy. The electron-channel measurement would be an obvious culprit for this discrepancy, because of bremsstrahlung and lower statistics with respect to the muon channel. On the other hand, disagreement is rather in the muon channel \cite{Aaij:2014pli,Aaij:2012vr}. A systematic effect there is less likely than in the electron channel, as muons are among the most reliable objects within LHCb.

The other mentioned $b \to s \mumu$ modes turn out to fit a coherent picture with $R_K$:
\begin{itemize}
\item The very same pattern, with data lower than the SM prediction, is also observed in the $B_{s} \rightarrow \phi \mu^{+} \mu^{-}$ channel and in the same range $m_{\mu \mu}^{2} \in [1, 6]$ GeV$^{2}$, as initially found in 1/fb of LHCb data \cite{Aaij:2013aln} and then confirmed by a full Run-1 analysis \cite{Aaij:2015esa}. This discrepancy is estimated to be more than 3$\sigma$ \cite{Aaij:2015esa}.
\item Additional support comes from the $B \rightarrow K^{*} \mu \mu$ angular analysis, exhibiting a discrepancy in one combination of the angular-expansion coefficients, known as $P_{5}'$.
\end{itemize}

\noindent The last point, known as ``the $P_{5}'$ anomaly'' deserves some further comments. From LHCb's full angular analysis of the decay products in $B \to K^* \mumu$, one can construct observables with limited sensitivity to form factors \cite{Descotes-Genon:2013vna}. One of such ``clean'' observables is called $P_{5}'$ as mentioned. The latter exhibits a discrepancy in two bins, again in the low-$m^{2}_{\mu\mu}$ range. It should (and has been, in the literature) stated clearly that this observable ought to be taken {\em cum grano salis}. In fact, what cancels is the dependence on the infinite-$m_b$ form factors. The crucial issue is how important departures from the infinite-$m_b$ limit are as the di-lepton invariant mass squared $q^2$ approaches the charmonium threshold $4 m_c^2$ -- in particular, departures due to $c \bar c$-loop contributions, that at present are still incalculable. While being formally power suppressed in a $1/m_b$ expansion, such $c \bar c$ contributions also come with a factor of $1 / (q^2 - 4 m_c^2)$ \cite{Khodjamirian:2010vf}, that becomes larger and larger as $q^2$ approaches the charmonium threshold. Because of this reason, it is difficult to estimate the actual significance of the $P_{5}'$ discrepancy, which is indeed rather debated, see in particular \cite{Descotes-Genon:2013wba,Lyon:2014hpa,Jager:2014rwa,Ciuchini:2015qxb}.

It is fair to say, however, that the $P_{5}'$ anomaly remains intriguing, because it occurs, again, in the same low-$q^2$ region of [1, 6] GeV$^2$, and because the effect, originally found in 1/fb of LHCb data \cite{Aaij:2013qta}, was confirmed by a full Run-1 analysis \cite{Aaij:2015oid} as well as, very recently, by a Belle analysis \cite{Abdesselam:2016llu}. From the experimental papers, the discrepancy amounts to $3.4\sigma$ as estimated by LHCb, whereas it is in the 2$\sigma$-ballpark from Belle (2.1$\sigma$ as compared to \cite{Descotes-Genon:2014uoa} and 1.7$\sigma$ as compared to \cite{Straub:2015ica,Jager:2012uw,Jager:2014rwa}).

Further interesting results come from measurements of the ratios $R(D^{(*)}) \equiv \mc B (B \to D^{(*)} \tau \nu) / \mc B (B \to D^{(*)} \ell \nu)$. They were initially reported by BaBar \cite{Lees:2012xj} to be in excess of the SM prediction \cite{Fajfer:2012vx,Kamenik:2008tj}. The tendency in the $R(D^{*})$ channel was recently confirmed by LHCb in 3/fb of Run-1 data \cite{Aaij:2015yra}. Consistent results were also reported by Belle in two analyses, using respectively hadronically- \cite{Huschle:2015rga} and semileptonically-decaying \cite{Belle-semilep} taus. A simultaneous fit to all these $R(D)$ and $R(D^*)$ measurements yields a discrepancy from the SM point with a significance exceeding 4$\sigma$ \cite{Belle-ICHEP16}.

\section{Theory considerations}
\label{sec:LFVth}

\noindent We can summarize the above experimental facts as follows. $R_K$ hints at Lepton Universality Violation (LUV), the effect being in muons rather than electrons. $R(D)$ and $R(D^*)$ also point to LUV, with in principle even larger statistical significance, although history teaches us we should stay prudent with measurements involving final-state taus. In this respect, the $R_K$ measurement is more solid, as it involves only light leptons. However, the $R_K$ significance is clearly still too low. Nonetheless, it is interesting that a whole range of other $b \to s \mumu$ modes display discrepancies that form a coherent picture with $R_K$.

In short, it is apparent that each of the mentioned effects needs confirmation from LHCb's Run 2 to be taken seriously. Yet, focusing for the moment on the $b \to s$ discrepancies, we can ask ourselves two questions: whether we can (easily) make theoretical sense of the above data, and what are the most immediate signatures to expect in case the above discrepancies are real.

As concerns the first question -- whether we can easily make theoretical sense of the experimental anomalies -- the answer is yes, within an effective-theory framework. Consider in fact the following Hamiltonian:
\bea
\label{eq:HSMNP}
\hspace{-0.3cm}\mc H_{\rm SM+NP}(\bar b \to \bar s\mu^+\mu^-) = -\frac{4G_F}{\sqrt{2}}V_{tb}^* V_{ts}\frac{\alpha_{em}(m_b)}{4\pi} \times \nn \\
\left[\bar b_L \gamma^\lambda s_L \, \bar \mu \left(C_9^{(\mu)} \gamma_\lambda 
 + C_{10}^{(\mu)} \gamma_\lambda \gamma_5\right)\mu \right]  + {\rm H.c.}\,,
\eea
where the index ${(\mu)}$ indicates that the Wilson coefficients of the corresponding operators (denoted as $\mc O_9$ and $\mc O_{10}$) distinguish among different lepton flavours, as the Hamiltonian on the l.h.s. includes new-physics (NP) contributions as well. The SM contributions for these Wilson coefficients are flavour universal, and such that $C_9 \simeq - C_{10}$ at the $m_b$ scale, yielding (accidentally) an approximate $(V-A) \times (V-A)$ structure. Advocating the same structure also for the corrections to $C_{9}^{\rm SM}$ and $C_{10}^{\rm SM}$ --  in the $\mu$-channel only! -- turns out to account at one stroke for $R_K$ lower than 1, $\mc B(B \to K \mu \mu)$ (and $\mc B(B_s \to \mu \mu)$) data below predictions, and the $P_5'$ anomaly in $B \to K^* \mu \mu$ data. A fully quantitative test of this statement requires a global fit, see in particular \cite{Ghosh:2014awa,Altmannshofer:2014rta}. These analyses show that the by far most favourite solutions are either a negative new-physics (NP) contribution to $C_9$, with $C_{9,\rm NP}^{(\mu)} \sim - 30\% \, C_{9,\rm SM}$, or by NP in the SU(2)$_L$-invariant direction $C_{9,\rm NP}^{(\mu)} = - C_{10,\rm NP}^{(\mu)}$, with $C_{9,\rm NP}^{(\mu)} \sim - 12\% \, C_{9,\rm SM}$. Note that such a solution is approximately RGE-stable.

We conclude that all $b \to s$ data can be explained if $C_9^{(\ell)} \approx - C_{10}^{(\ell)}$ and $|C_{9,\rm NP}^{(\mu)}| \gg |C_{9,\rm NP}^{(e)}|$. As pointed out in \cite{Glashow:2014iga}, this pattern can be generated from a purely third-generation interaction of the kind
\be
\label{eq:HNP}
\mc H_{\rm NP} = G \, (\bar b_L' \gamma^\lambda b_L') \, (\bar \tau_L' \gamma_\lambda \tau_L')~,
\ee
with $G = 1 / \Lambda_{\rm NP}^2$ a new Fermi-like coupling, corresponding to a NP scale $\Lambda_{\rm NP}$ in the TeV ballpark. The interaction in eq. (\ref{eq:HNP}) is expected, e.g., in partial-compositeness frameworks. The prime on the fields indicates that they are in the ``gauge'' basis, i.e. that below the EWSB scale they need to be rotated to the mass eigenbasis by usual chiral unitary transformations of the form
\bea
&&b_L' \equiv (d_L')_3 = (U_L^d)_{3i} (d_L)_i~, \nn \\
&&\tau_L' \equiv (\ell_L')_3 = (U_L^\ell)_{3i} (\ell_L)_i~,
\eea
whereby the r.h.s. fields represent the mass eigenbasis. These rotations, in general, induce LUV and Lepton-Flavor Violation (LFV) effects alike \cite{Glashow:2014iga}. This is actually a rather general expectation. In fact, consider a new, LUV interaction introduced to explain $R_K$, and defined above the electroweak symmetry breaking (EWSB) scale. Such interaction may be of the kind $\bar \ell Z^\prime \ell$, with $Z^\prime$ a new vector boson, or $\bar \ell \phi q$, with $\phi$ a leptoquark. The question arises, in what basis are quarks and leptons in the above interaction. Generically, it is not the mass eigenbasis -- this basis does not yet even exist, as we are above the EWSB scale. Then, rotating the $q$ and $\ell$ fields to the mass eigenbasis generates LFV effects, although the initial interaction was introduced to produce only LUV ones.

With the above ingredients we can straightforwardly explain $b \to s$ data. In fact, recalling our full Hamiltonian eq. (\ref{eq:HSMNP}), and denoting $\kappa_{\rm SM} \equiv -\frac{4G_F}{\sqrt{2}}V_{tb}^* V_{ts}\frac{\alpha_{em}(m_b)}{4\pi}$ the Wilson-coefficient normalization factor within the SM, the shift to the $C^{(\mu)}_9$ Wilson coefficient becomes
\be
\kappa_{\rm SM} C_9^{(\mu)} = \kappa_{\rm SM} C_{9}^{\rm SM} + \frac{G}{2} (U^d_L)^*_{33} (U^d_L)_{32} |(U^\ell_L)_{32}|^2~.~~
\ee
For the shift on the r.h.s. to explain the $R_K$ discrepancy, one needs destructive interference between the SM and NP contributions to $C_9^{(\mu)}$. This occurs for $G (U^d_L)_{32} < 0$, assuming $(U^d_L)_{33} \approx 1$. On the other hand, in the $ee$-channel one has
\be
\kappa_{\rm SM} C_9^{(e)} = \kappa_{\rm SM} C_9^{\rm SM} + \frac{G}{2} (U^d_L)^*_{33} (U^d_L)_{32} |(U^\ell_L)_{31}|^2~,~~
\ee
whereby the last term on the r.h.s. is negligible by assumption, as $|(U^\ell_L)_{31}|^2 \ll |(U^\ell_L)_{32}|^2$.

So, in the above setup one would have
\be
\label{eq:RKmodel}
R_K \approx \frac{|C_9^{(\mu)}|^2 + |C_{10}^{(\mu)}|^2}{|C_9^{(e)}|^2 + |C_{10}^{(e)}|^2} \simeq \frac{2 |C_{10, \rm SM} + C_{10, \rm NP}^{(\mu)}|^2}{2 |C_{10, \rm SM}|^2}~,
\ee
where the factors of 2 on the r.h.s. are due to the contributions from $|C_9|$ and $|C_{10}|$ being equal by assumption. The above expression is approximate as, in particular, phase-space factors are slightly different between the muon and the electron channels. Note as well that
\be
\label{eq:0.77}
0.77 \pm 0.20 = \frac{\mc B(B_s \to \mu \mu)_{\rm exp}}{\mc B(B_s \to \mu \mu)_{\rm SM}} = \frac{\mc B(B_s \to \mu \mu)_{\rm SM + NP}}{\mc B(B_s \to \mu \mu)_{\rm SM}} = \frac{|C_{10, \rm SM} + C_{10, \rm NP}^{(\mu)}|^2}{|C_{10, \rm SM}|^2}~,
\ee
implying, within the model in Ref. \cite{Glashow:2014iga}, the correlations (see also \cite{Hiller:2014yaa})
\be
\frac{\mc B(B_s \to \mu \mu)_{\rm exp}}{\mc B(B_s \to \mu \mu)_{\rm SM}} \simeq R_K \simeq \frac{\mc B(B^+ \to K^+ \mu \mu)_{\rm exp}}{\mc B(B^+ \to K^+ \mu \mu)_{\rm SM}}~.
\ee
According to the above relation, the measurement-over-SM ratio for $\mc B(B_s \to \mu \mu)$ provides a proxy for $R_K$. This is one more good reason to pursue accuracy in the $\mc B(B_s \to \mu \mu)$ measurement. Provided that the central value on the l.h.s. of eq. (\ref{eq:0.77}) does not increase, this test will be a sensitive one already by the end of Run 2, as the $\mc B(B_s \to \mu \mu)$ total error (dominated by the experimental component) is anticipated to be around 10\% \cite{Bediaga:2012py}.

\section{Experimental signatures}
\label{sec:LFVexp}

\noindent From the argument made above it is clear that, if $R_K$ is signaling beyond-SM LUV, then we may expect measurable LFV as well. This expectation holds true barring further theoretical assumptions that prevent LFV in the presence of LUV. As a general rule, the two types of effects go hand in hand. Assuming the interaction (\ref{eq:HNP}), the amount of LUV pointed to by $R_K$ actually allows to quantify rather generally \cite{Glashow:2014iga} the expected amount of LFV. In fact, $R_K$ yields the ratio
\be
\rho_{\rm NP} = -0.159^{+0.060}_{-0.070}
\ee
between the NP and the SM+NP contribution to $C_9^{(\mu)}$. Then, for example,
\be
\label{eq:BKll}
\frac{\mc B (B \to K \ell_i^\pm \ell_j^\mp)}{\mc B (B^+ \to K^+ \mu^+ \mu^-)} ~\simeq~ 
2 \rho_{\rm NP}^2 
\frac{| (U^\ell_L)_{3i} |^2 | (U^\ell_L)_{3j} |^2}{|(U^\ell_L)_{32}|^4}~,~~
\ee
implying
\bea
\label{eq:BKlilj}
\mc B (B \to K \ell_i^\pm \ell_j^\mp) \simeq 5\% \, \cdot \, \mc B (B^+ \to K^+ \mu^+ \mu^-) \, \cdot \, \frac{| (U^\ell_L)_{3i} |^2 | (U^\ell_L)_{3j} |^2}{|(U^\ell_L)_{32}|^4} \simeq &&\nn \\
[0.1cm]
\simeq 2.2 \times 10^{-8} \, \cdot \, \frac{| (U^\ell_L)_{3i} |^2 | (U^\ell_L)_{3j} |^2}{|(U^\ell_L)_{32}|^4}~,\hspace{2.4cm}&&
\eea
where we used $\mc B (B^+ \to K^+ \mu^+ \mu^-) \simeq 4.3 \times 10^{-7}$ \cite{Aaij:2014pli}, and neglected all terms proportional to the different masses of the final-state leptons.\footnote{Because of this approximation, eqs. (\ref{eq:BKll})-(\ref{eq:BKlilj}) provide only crude estimates in the case of decays involving a $\tau$ lepton. However, this approximation does not change the argument of the present paragraph.} Eq. (\ref{eq:BKlilj}) tells us that LFV $B \to K$ decays are expected to be in the ballpark of $10^{-8}$ times an unknown factor involving $U_L^\ell$ matrix entries. In the $\ell_i \ell_j = e \mu$ case, this ratio reads $|(U_L^\ell)_{31} / (U_L^\ell)_{32}| \lesssim 3.7$ \cite{Glashow:2014iga}, implying that the $B \to K \mu e$ rate may be around $10^{-8}$, or much less if $|(U_L^\ell)_{31} / (U_L^\ell)_{32}| \ll 1$. The latter possibility would suggest $U_L^\ell$ entries that decrease in magnitude with the distance from the diagonal. But then one may expect the ratio $|(U_L^\ell)_{33} / (U_L^\ell)_{32}| > 1$, implying a $B \to K \mu \tau$ rate of O($10^{-8}$) or above! In short, assuming the interaction (\ref{eq:HNP}), one can hope that at least one LFV $B \to K$ decay rate be in the ballpark of $10^{-8}$ \cite{Glashow:2014iga}, which happens to be within reach at LHCb's Run 2. An entirely analogous reasoning applies for the purely leptonic modes $B_s \to \ell_i^\pm \ell_j^\mp$, that may well be within reach of LHCb during Run 2, if the $U$-matrix factor on the r.h.s. is of order unity (or larger!) for at least one LFV mode.\footnote{%
For a (rough) comparison, we should keep in mind that at Run 2 the LHCb is expected \cite{Bediaga:2012py} to provide a first measurement of $\mc B (B_d \to \mu^+ \mu^-)$, which in the SM is about 3\% of $\mc B (B_s \to \mu^+ \mu^-)$.}

It is worthwhile to open two parentheses on the consequences of the above argument.
First, it is an order-of-magnitude argument, and it is meaningful to speculate on the possibility of more quantitative LFV predictions. This possibility requires knowledge of the $U_L^\ell$ matrix. One approach towards predicting the $U_L^\ell$ matrix is the one pursued in Ref. \cite{Guadagnoli:2015nra}, whose line of argument goes as follows. A sufficient condition for $U_L^\ell$ to be predictable is to know the product $Y_\ell Y_\ell^\dagger$, with $Y_\ell$ the charged-lepton Yukawa coupling. To this end, one may start from the ansatz in \cite{Appelquist:2015mga} that the five flavour-SU(3) rotations are not all independent. Choosing three to be the independent ones allows to predict one SM Yukawa coupling in terms of the other two. One can thereby determine $Y_\ell$ in terms of $Y_u$ and $Y_d$. However, we don't know $Y_u$ and $Y_d$ in full. Yet, we can take an independently motivated model for $Y_u$ and $Y_d$ textures, such as the one in Ref. \cite{Martin:2004ec}, motivated as a scenario for solving the strong-CP problem in QCD. Another approach \cite{Boucenna:2015raa} starts from the observation that the product $(U_L^\ell)^\dagger U_L^\nu$ equals a known object, namely the PMNS matrix. Making assumptions about $U_L^\nu$ then allows to predict $U_L^\ell$. In this respect, Ref. \cite{Boucenna:2015raa} makes the ansatz $U_L^\nu = 1$.

A second parenthesis concerns the observation that the $B_s \to e \mu$ is expected to be the most difficult to access among the above-mentioned LFV modes, because it is chirally suppressed, and because the involved lepton combination is the farthest from the third one. It is therefore useful to search for additional decays that can give access to the same physics, while being comparably (or, hopefully, more) accessible experimentally. As pointed out in Ref. \cite{Guadagnoli:2016erb}, in the $B_s \to \mu e$ channel one such `proxy' decay is provided by the inclusion of an additional hard photon in the final state. In fact, the additional photon replaces the chiral-suppression factor, of order ${\rm max}(m_{\ell_1},m_{\ell_2})^2 / m_{B_s}^2$, with a factor of order $\alpha_{\rm em} / \pi$. The actual enhancement of $\mc B(B_s \to \mu e \gamma)$ is of about 30\% \cite{Guadagnoli:2016erb} over the non-radiative counterpart. Therefore, inclusion of the radiative mode would allow to more-than-double statistics with respect to the non-radiative mode alone.

The interaction advocated in eq. (\ref{eq:HNP}) has implications in $K$ physics as well, in decays of the kind $K \to (\pi) \ell \ell'$, such as $K_L \to e^\pm \mu^\mp$ and $K^+ \to \pi^+ e^\pm \mu^\mp$. Experimental limits on these modes are more than ten years old: $\mc B(K_L \to e^\pm \mu^\mp) < 4.7 \times 10^{-12}$ \cite{Ambrose:1998us}, $\mc B(K^+ \to \pi^+ e^- \mu^+) < 1.3 \times 10^{-11}$ \cite{Sher:2005sp}, $\mc B(K^+ \to \pi^+ e^+ \mu^-) < 5.2 \times 10^{-10}$ \cite{Appel:2000tc}. Theoretical expectations for the above decays are straightforwardly calculable after suitably normalising the decay modes of interest in order to cancel phase-space factors. Defining $\beta^{(K)}$ as the ratio of the new-physics Wilson coefficient responsible for the decay in the numerator over the SM Wilson coefficient responsible for the normalising decay, we obtain
\bea
\label{eq:LFV_K_decays}
&&\frac{\Gamma(K_L \to e^\pm \mu^\mp)}{\Gamma(K^+ \to \mu^+ \nu_\mu)} = | \beta^{(K)} |^2~, \\
\label{eq:LFV_K_decays_2}
&&\frac{\Gamma(K^+ \to \pi^+ \mu^\pm e^\mp)}{\Gamma(K^+ \to \pi^0 \mu^+ \nu_\mu)} = 4 | \beta^{(K)} |^2~.
\eea

To get a numerical idea of the effects to be expected, we need a model predicting $|\beta^{(K)}|^2$. For the sake of definiteness, here we use ``model A'' of Ref. \cite{Guadagnoli:2015nra} (any other motivated model, for example Ref. \cite{Boucenna:2015raa}, will do), thereby obtaining $| \beta^{(K)} |^2 = 2.15 \times 10^{-14}$. Use of eqs. (\ref{eq:LFV_K_decays}) then implies
\be
\label{eq:KLemu}
\mc B(K_L \to e^\pm \mu^\mp) \approx 6 \times 10^{-14}~,
\ee
where we have taken $\mc B(K^+ \to \mu^+ \nu_\mu) \approx 64\%$ and $\Gamma(K^+) / \Gamma(K_L) \approx 4.2$ \cite{Agashe:2014kda}. In addition
\be
\mc B(K^+ \to \pi^+ e^\pm \mu^\mp) \approx 3 \times 10^{-15}
\ee
after use of $\mc B(K^+ \to \pi^0 \mu^+ \nu_\mu) \approx 3\%$.

While the $K^+$ LFV mode is clearly too suppressed (within the considered model!), the $K_L$ one, eq. (\ref{eq:KLemu}), has a branching ratio close to $10^{-13}$. Such a rate may actually be reachable at the NA62 experiment. As concerns LHCb, it should be noted that, although $K$ mesons are produced copiously, their lifetimes are typically too long for the detector size -- with the exception of the $K_S$. A dedicated study is thus necessary to understand the actual LHCb capabilities for the above decays.

\section{More signatures}
\label{sec:more}

Being defined above the EWSB scale, our assumed operator, eq. (\ref{eq:HNP}), must actually be made invariant under the full SM gauge group \cite{Alonso:2014csa}. This operation yields interactions of the kind $(\bar Q_L^{\prime i} \gamma^\lambda Q_L^{\prime i})\,(\bar L_L^{\prime j} \gamma_\lambda L_L^{\prime j})$ and $(\bar Q_L^{\prime i} \gamma^\lambda Q_L^{\prime j})\,(\bar L_L^{\prime j} \gamma_\lambda L_L^{\prime i})$, with $i, j$ SU(2)$_L$ indices and $Q_L^\prime$, $L_L^\prime$ the SM quark and lepton doublets in the gauge basis. The second interaction yields in turn charged currents like $(\bar t^\prime_L \gamma^\lambda b_L^\prime)(\bar \tau^\prime_L \gamma_\lambda \nu_{\tau L}^\prime)$. After rotation to the mass eigenbasis, the last structure contributes to $\Gamma(b \to c \tau \nu_\tau)$ \cite{Bhattacharya:2014wla}, thereby allowing to explain the LHCb and $B$-factories deviations on $R(D^{(*)})$.

However, this coin has a flip side. Properly taking into account renormalization-group running from the NP scale to the relevant low-energy scale, one finds non-trivial constraints \cite{Feruglio:2016gvd}, in particular from $B \to K \bar \nu \nu$ (see also \cite{Calibbi:2015kma}), from LEP-measured $Z \to \ell \ell$ couplings, and especially from LUV effects in $\tau \to \ell \nu \nu$ decays. The latter constraints are the most dangerous, as they are tested to per mil accuracy, and they turn out to ``strongly disfavor an explanation of the $R(D^{(*)})$ anomaly  model-independently'' \cite{Feruglio:2016gvd}. The same argument shows that also LFV decays of leptons are generated, and that they provide probes well competitive with the ones pointed out above, in particular $\mc B(\tau \to 3\mu)$, $\mc B(\tau \to \mu \rho) \sim 5 \times 10^{-8}$ \cite{Feruglio:2016gvd}.

\section{Model-building considerations}
\label{sec:models}

\noindent Up to this point, we have limited ourselves to Effective-Field-Theory (EFT) considerations, we have namely discussed Fermi-like interactions involving SM fields only. An immediate question is whether any plausible dynamics able to generate these EFT interactions exists. A positive answer to this question has to face a few challenging obstacles. The first one, relevant if we seek a common explanation of $b \to s \ell \ell$ and $b \to c \tau \nu$ discrepancies, is the fact that, on the one hand, $B \to D^{(*)} \tau \nu$ arises at tree level in the SM, and the effect is as large as O(25\%). This would call for tree-level charged mediators. On the other hand, the effects in $B \to K^{(*)} \ell \ell$ decays, while being again $\lesssim 25\%$, are corrections to processes that in the SM arise at {\em loop} level.

A second obstacle is inherent in the fact that the needed NP is of the kind $J_{\rm quark} \times J_{\rm lepton}$, i.e. as the product of a quark and a lepton current. It is hard to believe that such NP leaves no traces in processes of the kind $J_{\rm quark} \times J_{\rm quark}$ and $J_{\rm lepton} \times J_{\rm lepton}$ as well. To the former category belong notably $B_s$-mixing observables, and to the latter purely leptonic LFV or LUV decays. Both classes of observables pose, in general, formidable constraints.

Finally, a third obstacle is evident from the observation that most (all?) model-building attempts that have been made involve
\begin{itemize}
\item new charged, and possibly colored, states, 
\item with masses in the TeV region ($R_K$ and $R(D^{(*)})$ effects are large, so the new states cannot be too heavy) and
\item with significant couplings to 3$^{\rm rd}$-generation SM fermions.
\end{itemize}
From these conditions one may expect that constraints from direct searches be potentially strong. And indeed they are! Searches of resonances decaying to $\tau \tau$ pairs are of special relevance, see \cite{Greljo:2015mma,Faroughy:2016osc}.

All of the above being said, many attempts towards plausible UV completions able to produce the needed EFT operators have been made. The proposed models involve typically the introduction of
a new Lorentz-scalar or -vector, with any \footnote{Of course, here `any' should be understood as all those allowed by gauge invariance.} of the following transformation properties under the SM gauge group:
\begin{itemize}
\item SU(3)$_c$: a singlet or a triplet, the latter case referred to in the following as a leptoquark,
\item SU(2)$_L$: a singlet or a doublet or a triplet.
\end{itemize}
In the following I will shortly review only the models proposed to account simultaneously for the $b \to s$ and the $b \to c$ anomalies, with the main aim of exposing the non-trivial challenges that such simultaneous explanation poses in the face of all the existing constraints.

\medskip

{\bf Explicit Models --} Among the combinations mentioned in the previous paragraph, a first natural possibility is that of a color-singlet, weak-triplet vector field, i.e. a heavier replica of the $W^\pm, Z^0$ bosons. This possibility is discussed in Refs. \cite{Greljo:2015mma,Boucenna:2016wpr}. In particular, Ref. \cite{Greljo:2015mma} studies the strong bounds coming from $\tau \to \ell \nu \nu$ as well as $B_s$-mixing, and also confronts the model with direct searches. The conclusion is that the minimal model is indeed ruled out by searches of resonances decaying to $\tau$ pairs. This constraint can be circumvented by advocating less minimal versions of the model, at the cost of introducing more free parameters. Ref. \cite{Boucenna:2016wpr} considers in full generality the possibility of gauge extensions that lead to LUV and may thus address the flavour anomalies. It concludes that the most promising candidate models (in the light of a number of inescapable constraints) are gauge extensions whereby gauge couplings are universal, whereas non-universality arises from Yukawa couplings of the SM fermions with new vector-like fermions. I do not discuss in detail the other color-neutral possibilities (weak triplet of scalars, and weak doublets or singlets). They either amount to extended Higgs sectors, that in general have a tree-level FCNC problem, or else fail to fulfil gauge invariance.

More numerous model-building attempts exist for color-triplet scalars or vectors. As mentioned before, color triplets are usually referred to as leptoquarks (LQ) \cite{Davidson:1993qk}, namely states coupled to a quark and a lepton. One reason for their appeal is the fact that they are not subject to constraints from meson mixings.\footnote{In addition, Lorentz-vector LQ states naturally appear in grand-unified theory (GUT) scenarios, although in this case one would expect their mass to lie close to the GUT scale, whereas the flavour anomalies would require a much lower scale.} A first proposal of a LQ model able to explain both $R_K$ and $R(D^{(*)})$ is Ref. \cite{Bauer:2015knc}. While the proposed dynamics is rather convincing -- in that corrections to $R(D^{(*)})$ arise at tree level, whereas $R_K$ is only generated at one loop -- this model turns out not to be viable, see discussion in Ref. \cite{Becirevic:2016oho}. A model-building attempt in the direction of a weak-triplet Lorentz-vector with completely general flavour couplings $g_{ij}$ to a $\bar Q^i_L L^j_L$ bilinear is presented is Ref. \cite{Fajfer:2015ycq}, $i,j$ denoting flavour indices. (This scenario generalises Ref. \cite{Calibbi:2015kma}, where the LQ was assumed to couple only to the third-generation fermions in the weak basis.) These fully general flavour couplings are `pragmatically' fit to data. Indicating with $M_U$ the mass of the LQ, the analysis returns constraints of the kind $g_{b \mu}^* g_{s \mu} \sim 10^{-3} \cdot (M_U/\mbox{TeV})^2$ and $|g_{b \tau}| \gtrsim 2$ from $b \to s$ and respectively $b \to c$ anomalies. One may argue that this hierarchy introduces another flavour problem. Besides, the very presence of a LQ mass begs for a mechanism generating it, while preserving gauge invariance.

A rather extensive LQ study was presented in Ref. \cite{Barbieri:2015yvd}, encompassing the cases of a weak-singlet vector, and of a weak-singlet scalar or vector. The paper starts from the observation that $b \to c \tau \nu$ anomalies are a deviation from a SM tree amplitude involving the third generation of leptons, whereas the $b \to s \ell \ell$ ones are corrections to a SM loop amplitude with the light generations of leptons only. In the light of these differences, the authors speculate whether a flavour group $G_F$ and a tree-level LQ-exchange mechanism exist such that: {\em (i)} in the limit of exact $G_F$, the LQ couples only to the third generation of SM fermions, and {\em (ii)} the needed NP effects arise from the $G_F$ breaking, giving rise predominantly to corrections to $b \to c \tau \nu$ and, at a weaker level, to effects in $b \to s \ell \ell$ as well. This approach is again fairly convincing, as it makes sense of all the anomalies from the controlled breaking of a global (flavour) symmetry. However, since the only $G_F$-invariant SM fermions are the left-handed doublets, the generated EFT operators will not escape the general argument in Ref. \cite{Feruglio:2016gvd} presented above. On the other hand, a merit of Ref. \cite{Barbieri:2015yvd} is that it exposes the strong UV-cutoff sensitivity of LQ  models involving a Lorentz vector, sensitivity manifest in the power-like divergence of 2-, 3-point functions, and box diagrams.\footnote{The latter reintroduce at one loop the constraint from $B$-meson mixings, that for LQs is absent at tree level, as mentioned above.} In the light of such cutoff sensitivity, any of these EFT models badly needs -- even for just the sake of calculability -- an explicit UV completion. A separate challenge is then represented by the detailed verification that this UV completion does withstand all the existing constraints.

Finally, a recent example of a simple viable model is Ref. \cite{Becirevic:2016yqi}, advocating a weak-doublet scalar LQ coupled to the bilinears $\bar d_R L_L$ and $\bar Q_L \nu_R$ through $Y_L$ and $Y_R$ Yukawa couplings. (Notice the presence of the right-handed $\nu$ field, required to have negligibly small mass.) By virtue of the $(V + A)_{\rm quark} \times (V-A)_{\rm lepton}$ current invoked, this setup is not affected by the constraint in Ref. \cite{Feruglio:2016gvd}, and predicts $R_{K^*} > 1$ \cite{Hiller:2014ula}, which awaits the verdict of LHCb's Run~2.

\section{Further tests}
\label{sec:tests}

\noindent The overall conclusion of the above theory discussion is that, while all the flavour anomalies are neatly fit at the EFT level, less straightforward and compelling are the attempts towards UV dynamics responsible for these effective interactions. The bright side of the story is that a whole range of tests has been proposed in the literature to clarify the situation, in the first place experimentally, and that most if not all of these tests are within reach at Run 2 of the LHCb experiment, not to mention the upcoming Belle-2 startup. I classify these tests into three broad categories based on personal judgment:
\begin{itemize}
\item Measurements of additional LUV ratios;
\item Extraction of long-distance effects from {\em data};
\item Definition and measurement of new observables sensitive to $C_9$ and $C_{10}$.
\end{itemize}
In the following I will briefly discuss each of these directions in turn.

The first direction is rather evident: a first strategy towards ascertaining the reality of LUV is (besides more $R_K$ and $R(D^{(*)})$ determinations) to measure LUV ratios in different channels. In a notation generalising eq. (\ref{eq:RK}), one may measure $R_{K^*}$, $R_\phi$, $R_{K_0(1430)}$, $R_{f_0}$, and the inclusive $R_{X_s}$. The quantitative channel-by-channel LHCb reach is as yet not fully clear, although an $R_{K^*}$ measurement is imminent at the time of this writing. An interesting test \cite{Hiller:2014ula} is to define the double ratios
\be
X_H \equiv \frac{R_H}{R_K}~,
\ee
with $H = K^*,$ $\phi$, $K_0(1430)$, $f_0$ or $X_s$. Deviations from unity in the double ratios $X_H$ can only come from right-handed quark currents.

Especially in $b \to s \ell \ell$ data, an important obstacle towards a robust comparison of data with theory is the presence of long-distance (LD) effects due to $c \bar c$ loops. As discussed, these effects escape at present a first-principle calculation and increase in importance as the di-lepton invariant mass squared approaches the charmonium threshold ($q^2 \simeq 4 m_c^2$). In this respect, it is unclear whether such effects may spill over and pollute the low-$q^2$ region [1, 6] GeV$^2$ \cite{Lyon:2014hpa}, to an extent able to explain away the $b \to s \ell \ell$ anomalies. Note on the other hand that, while such argument may work for anomalies involving absolute branching ratios, it falls short of explaining away ratios such as $R_K$. Encouraging is the fact that such matter seems amenable to be sorted out experimentally. One may measure the $m_{\mumu}$ spectrum, including the $c \bar c$ resonances as a sum of Breit-Wigner shapes weighed by {\em complex} coefficients (thus even accounting for interference effects), and fit this parameterization {\em to data} \cite{Lyon:2014hpa,Kruger:1996cv}. A first example of such an approach was recently presented in Ref. \cite{Aaij:2016cbx}. By this method the LHCb collaboration performed a new measurement of the $B^+ \to K^+ \mumu$ branching ratio across the full $q^2$ range. It is reassuring that the measurement yields a result compatible with previous measurements \cite{Aaij:2014pli}, and intriguing that, again, the result is {\em below} the SM prediction \cite{Bailey:2015dka}.

A third, crucial class of tests towards establishing the existing flavour anomalies is to devise and measure new observables, independently sensitive to $C_9$ and $C_{10}$. A possible example in this sense is offered by the $B_s \to \mumu \gamma$ decay. Its branching ratio is, for low $q^2$, sensitive to the Wilson coefficient $C_7$ of the electromagnetic-dipole operator, and, in the whole $q^2$ range, to the $C_{9,10}$ Wilson coefficients and their right-handed counterparts. Furthermore, its total branching ratio is one order of magnitude above the $B_s \to \mumu$ one, because the chiral suppression in the latter decay is replaced by an $\alpha_{\rm em} / \pi$ factor. However, a direct measurement of the $B_s \to \mumu$ decay -- namely a measurement aiming at detecting and reconstructing the radiated photon -- poses a major challenge at hadron colliders. In view of this limitation, Ref. \cite{Dettori:2016zff} proposed a novel method, whereby the $B_s \to \mumu \gamma$ decay is searched for in the very same event sample selected for the $\mc B(B_s \to \mumu)$ measurement. The idea is to access $B_s \to \mumu \gamma$ as ``contamination'' to $B_s \to \mumu$ as the signal window for the latter search is suitably enlarged downwards. This approach makes sense because the initial-state-radiation (ISR) and final-state-radiation (FSR) components are relevant in different $q^2$ regions of the $B_s \to \mumu \gamma$ differential decay width ($q^2$ being the di-lepton invariant mass squared), and ISR-FSR interference is negligibly small in the whole $q^2$ range -- this holding true in any plausible SM extension. As a consequence, the ISR and FSR components in $\mc B(B_s \to \mumu \gamma)$ can be treated as independent \cite{Dettori:2016zff}. Furthermore, the FSR component can be systematically subtracted from data, the same way it is in the $B_s \to \mumu$ decay measurement.\footnote{As well known in fact, the decay $B_s \to \mumu + n \gamma$ with $n \gamma$ denoting an arbitrary number of soft, undetected, bremsstrahlung photons, inevitably pollutes the $B_s \to \mumu$ decay measurement as $q^2$ approaches the $m_{B_s}^2$ peak \cite{Buras:2012ru}. As such, this component needs to be quantified through a Monte Carlo \cite{Golonka:2005pn}, and accounted for by event reweighting.} Therefore, the strategy proposed in Ref. \cite{Dettori:2016zff} gives access to $B_s \to \mumu \gamma$ in a $q^2$ region where the decay is completely dominated by the ISR component of the photon spectrum. This measurement can be compared with the SM prediction, as computed in Ref. \cite{Melikhov:2004mk} to leading order in $\alpha_{\rm em}$ (see also Ref. \cite{Kozachuk:2016ypz}). The by far dominant source of uncertainty in this calculation comes from the $B_s \to \gamma$ vector and axial form factors, for which no first-principle calculation exists. The form-factor predictions are obtained from \cite{Kozachuk:2015kos}, an analysis based on the relativistic constituent quark model \cite{Anisovich:1996hh,Melikhov:2001zv}, tested to reproduce the known results from QCD for heavy-to-heavy and heavy-to-light form factors \cite{Melikhov:2000yu}. The predictions used in Ref. \cite{Dettori:2016zff} are thereby attached a 20\% uncertainty, clearly not yet sufficient to fully resolve the effects expected from NP. On the other hand, what is required for the proposed method are the form factors in the high-$q^2$ range close to the kinematic endpoint. This range is the preferred one for lattice-QCD simulations. This method can realistically be applicable in LHC Run 2 data, and would allow to set the first limit for $\mc B (B_s \to \mumu \gamma)$, or provide the first measurement thereof.

\section{Conclusions}
\label{sec:conclusions}

\noindent In flavour physics there are by now several persistent discrepancies with respect to the SM. Their most convincing aspects are the following: experimentally, results are consistent between LHCb and $B$ factories; deviations concern two independent sets of data, namely $b \to s$ and $b \to c$ decays; discrepancies go in a consistent direction and a beyond-SM explanation is possible already within an effective-theory approach. It is of course far too early to draw conclusions, as the above effects await confirmation from Run 2 -- but the verdict will come very soon.

The above situation suggests the following remarks as an outlook. At the theoretical level, while as mentioned data fit coherently an effective-theory picture, it is hard to find convincing UV dynamics that produces the required effective operators {\em and} withstands all existing constraints. Therefore, more model-building may have to wait for more experimental information. Conversely, at the experimental level it is timely to pursue further tests to get more insights. Examples of these tests include: more measurements of $R_K$ and of other LUV ratios; estimates of long-distance $c \bar c$-loop effects directly from data; and finally, the proposal and measurement of more observables sensitive to the relevant Wilson coefficients $C_9$ and $C_{10}$.

The topic reviewed here illustrates the enormous richness of flavour data, and their potential to better understand SM dynamics, including non-perturbative QCD, and to uncover new dynamics, even if it is out of reach for direct searches.

\section*{Acknowledgments}

I thank my collaborators of refs. \cite{Glashow:2014iga,Guadagnoli:2015nra,Guadagnoli:2016erb,Dettori:2016zff} for fruitful and enjoyable discussions. This work is partially supported by the CNRS grant PICS07229.


\end{document}